\documentclass[lettersize,journal]{IEEEtran}
\usepackage{amsmath,amsfonts}
\usepackage{algorithmic}
\usepackage{algorithm}
\usepackage{array}
\usepackage[caption=false,font=normalsize,labelfont=sf,textfont=sf]{subfig}
\usepackage{textcomp}
\usepackage{stfloats}
\usepackage{url}
\usepackage{verbatim}
\usepackage{graphicx}
\usepackage{cite}
\usepackage{hyperref}

\begin{document}

\title{Decoupled Delay Compensation: Enhancing Pre-trained MARL Policies via Learned Dynamics Filtering}

\author{Maxim Mednikov, Oren Gal
\thanks{All authors are with the University of Haifa, Israel}
}

\markboth{}{ \MakeLowercase{\textit{et al.}}: execution belief-state filtering layer}


\maketitle

\begin{abstract}
Real-world multi-agent reinforcement learning (MARL) systems must often operate under stale observations, stochastic communication delays, and intermittent packet loss. Policies trained under idealized synchronous conditions frequently exhibit significant performance degradation in these regimes because they act on outdated feedback. We propose a modular execution-stage state-estimation layer that replaces delayed communicated observations with current belief-state estimates. The framework integrates a learned Gated transition model with a recursive Kalman filtering layer to estimate instantaneous states from asynchronous measurements. A primary advantage of this approach is its modularity, The estimator serves as a plug-in for pre-trained policies, requiring no modifications to the original MARL training algorithm, architecture, or reward structure. Evaluation across diverse multi-agent and continuous-control benchmarks demonstrates that the proposed layer consistently enhances robustness to communication latency and message loss. The most significant performance gains are observed in coordination-intensive and dynamically unstable tasks where temporal consistency is critical for control.
\end{abstract}


\section{Introduction}
\IEEEPARstart{M}{ulti-Agent} Reinforcement Learning (MARL) has achieved significant success in domains ranging from strategic games to coordinated robotic control. However, as the number of agents increases, the dimensionality of the joint action-observation space grows, further complicating the credit assignment problem and the optimization of free parameters \cite{gronauer_multi-agent_2022}. Because MARL algorithms are predominantly trained in simulation, they often encounter a significant Sim2Real gap during physical deployment \cite{da_survey_2025}. A critical component of this gap, and the focus of this work, is the reliance on idealized execution models where observations are perfectly synchronized, communication is lossless, and actions are instantaneous. Physical systems seldom meet these criteria, as they must operate under stochastic transmission delays, intermittent packet loss, and sensor-processing latency \cite{nath_revisiting_2021}. When policies trained under idealized conditions encounter stale information, performance often degrades sharply, leading to control instability \cite{neto_reinforcement_2026}.

This discrepancy is significant because decentralized policies are highly sensitive to temporal consistency; delays effectively violate the decentralized Markov assumption \cite{katsikopoulos_markov_2003}. A message received from a peer may describe a historical rather than instantaneous state, forcing the agent to operate on a non-causal observation stream. This temporal misalignment introduces spurious non-stationarity, as the perceived state configuration no longer accurately represents the immediate environment. Existing mitigation strategies primarily involve training-time interventions, such as state augmentation or specialized delay-aware architectures. While effective, these methods require access to policy internals, necessitate end-to-end retraining for specific delay regimes, and can exacerbate the curse of dimensionality \cite{hernandez-leal_survey_2017}. Such requirements limit their applicability to expensive-to-train models, legacy systems, or proprietary controllers where parameter modification is not feasible.

In this work, we treat communication delay as an execution-stage state-estimation problem. Instead of retraining the policy, we introduce a belief-state filtering layer between the environment and a pre-trained decentralized controller. This layer replaces delayed observations with current estimates of task-relevant state variables. Specifically, we integrate a learned Gated Recurrent Unit (GRU) transition model \cite{chung_empirical_2014} into a recursive Kalman structure \cite{noauthor_seminal_nodate}. The estimator predicts state transitions through missing or delayed packets using a residual dynamics formulation and assimilates asynchronous information as it arrives, providing a synchronized belief representation for action selection.

The proposed framework is modular by design. Because the filtering layer operates exclusively at execution time, it can be integrated with existing MARL policies without altering the underlying learning algorithm, critic, or training objective. This decoupling is particularly advantageous for heterogeneous robot fleets where retraining individual agent policies to match varying hardware latencies is computationally prohibitive. We evaluate this approach across discrete and continuous multi-agent benchmarks with varying observability and simulation fidelity. Results demonstrate that the layer consistently enhances robustness to delay and packet loss, particularly in tasks sensitive to high-frequency feedback. Furthermore, we show that the recursive structure mitigates observation noise common in real-world sensor deployments.

The main contributions of this paper are:
\begin{itemize}
\item \textbf{A Modular Execution Framework:} We formulate delayed multi-agent communication as a recursive belief-state estimation problem, enabling delay-robustness for pre-trained policies without the need for retraining or policy modification.
\item \textbf{Learned-Recursive Estimation Layer:} We introduce a plug-in module combining a GRU-based world model with a Kalman filtering update to handle stochastic delay and intermittent packet loss.
\item \textbf{Cross-Domain Empirical Validation:} We demonstrate that this approach generalizes across diverse MARL backbones and dynamical systems, Preserving policy performance even under out-of-distribution delay patterns.
\end{itemize}

\section{Preliminaries}
\subsection{Belief space}
In partially observable settings, an agent can maintain a \emph{belief state} $b_t \in \Delta S$, i.e., a posterior distribution over the latent state $s_t$ given its history $h_t=(o_{0:t},a_{0:t-1})$:
\begin{equation}
    b_t(s) := P(s_t=s \mid h_t).
\end{equation}
Given an action $a_{t-1}$ and a new observation $o_t$, the belief evolves via the Bayesian filtering recursion (``predict'' then ``update''):
\begin{align}
    \bar{b}_t(s') &= \sum_{s \in S} P(s' \mid s, a_{t-1})\, b_{t-1}(s), \\
    b_t(s') &= \eta\, P(o_t \mid s')\, \bar{b}_t(s'),
\end{align}
where $\eta$ normalizes $b_t$. Importantly, this recursion induces a Markovian dynamics in belief space, enabling policies to be written as $\pi(a_t\mid b_t)$ instead of depending on the full history.

Exact belief tracking is typically intractable for continuous, high-dimensional systems \cite{bongard_probabilistic_2008}. Under linear dynamics with Gaussian process/measurement noise, the belief is fully characterized by its mean and covariance, yielding the classic \emph{Kalman Filter \cite{noauthor_seminal_nodate}}.

\subsection{Multi-Agent Reinforcement Learning}
Multi-Agent Reinforcement Learning (MARL) involves n agents interacting within a shared environment, where state transitions and individual rewards depend on the \emph{joint} action. This framework is typically formalized as a stochastic Markov game defined by a global state $s_t$, per-agent observations $o_{i,t}$ and actions $a_{i,t}$. The joint action $\mathbf{a}_t=(a_{1,t},\dots,a_{n,t})$ drives the transition model $P(s_{t+1}\mid s_t,a_t)$ and per-agent rewards $r_{i,t}$.

In practical systems, agents often operate undependable using only local and communicated information, leading to partial observability and non-stationarity as neighboring policies evolve during the training process. A common paradigm to address this is \emph{centralized training with decentralized execution} (CTDE): global information is utilized during training (e.g., by a centralized critic), while agents rely on decentralized policies during execution.

In the considered delayed-observation setting, agent $i$ conditions its policy on a belief state $b_{i,t}$ rather than raw observations. The objective is to optimize decentralized policies $\{\pi_{\theta_i}\}_{i=1}^n$ to maximize the expected discounted return:
\begin{equation}
    J(\theta) = \mathbb{E}\left[\sum_{t=0}^{\infty} \gamma^t r_t(s_t,\mathbf{a}_t)\right].
\end{equation}

\subsection{Decentralized partially observable Markov decision process}
We formalize the cooperative multi-agent setting as a decentralized partially observable Markov decision process (Dec-POMDP), defined by the tuple $\langle S, \{A_i\}_{i=1}^n, P, R, \{O_i\}_{i=1}^n, \{\omega_i\}_{i=1}^n, \gamma, \mu \rangle$. Here, $S$ is the latent global state, each agent $i$ selects an action $a_{i,t}\in A_i$, and the joint action $\mathbf{a}_t=(a_{1,t},\dots,a_{n,t})$ drives transitions $s_{t+1}\sim P(s_t,\mathbf{a}_t)$. Each agent receives only a local observation $o_{i,t}\sim \omega_i(s_t)$ and a shared team reward signal in the cooperative case.

The key challenge is that each agent must choose actions from its own local history $h_{i,t}=(o_{i,0:t},a_{i,0:t-1})$, whereas environmental dynamics depend on the full joint process. This difficulty is compounded by delayed or intermittent communication, which results in stale or inconsistent observations across agents \cite{nath_revisiting_2021,fu_rainbow_2025}.

Common approaches to this problem involve state augmentation or recurrent neural networks to encode histories into latent representations \cite{hausknecht_deep_2017,heess_memory-based_2015}. In contrast, we operate in \emph{belief space}, where each agent maintains a belief $b_{i,t}$ over control-relevant latent variables. This provides a compact state estimate derived from past information, facilitating robust execution under communication delays.

\section{Related Work}

\subsection{Delays in RL and MARL.}
In delayed environments, the observations accessible to the policy at time $t$ often fail to reflect the instantaneous state, instead containing stale information due to sensor or actuation latency. A standard mitigation strategy is \emph{state augmentation}, which provides the agent with a fixed window of historical observations and actions to recast the delayed execution as an augmented Markov process \cite{chen_delay-aware_2020}. While effective for deterministic delays, this approach scales poorly when delays are stochastic, significantly large, or coupled with observation loss \cite{nath_revisiting_2021}. Consequently, recent studies have investigated explicit latency modeling, such as value functions that directly incorporate delay uncertainty \cite{yu_delay-aware_2024}.

In MARL, these challenges are compounded by inter-agent communication constraints, where agents must maintain coordination while operating on stale or asynchronous messages under partial observability. Prior work has primarily addressed this by optimizing communication schedules or introducing training-time compensation mechanisms. For example, delay-aware communication modules adapt exchange timing to balance responsiveness with coordination quality \cite{yuan_dacom_2022}. Similarly, Rainbow Delay Compensation (RDC) \cite{fu_rainbow_2025} integrates learned compensation, curriculum learning, and knowledge distillation into the MARL training pipeline.

Our work is complementary to these approaches. Rather than proposing an alternative delay-aware training algorithm, we focus on the execution stage: we preserve the original policy backbone and insert a state-estimation layer that transforms delayed communication into a current belief-state estimate during inference.

\subsection{Filtering and learned state estimation.}
Another research direction treats delay-induced uncertainty as a \emph{state-estimation} problem. Under this paradigm, the primary objective is to maintain a belief over the latent state rather than processing an augmented history. While linear-Gaussian systems utilize standard Kalman filtering, nonlinear dynamics require approximations such as Extended (EKF) or Unscented Kalman Filter (UKF) variants. Furthermore, hybrid architectures like KalmanNet integrate neural networks with classical recursive structures to mitigate model mismatch and handle complex nonlinearities \cite{revach_kalmannet_2022,revach_unsupervised_2021}.

State-estimation approaches have demonstrated significant efficacy in delayed single-agent control, with applications ranging from autonomous drone navigation using IMM-Kalman prediction \cite{marino_beyond_2024} to model-based RL for robotics under stochastic observation delays \cite{karamzade_model-based_2025}. Complementary methods focus on learning compact latent belief representations to mitigate partial observability \cite{wang_learning_2023} or employing stochastic cloning to integrate out-of-sequence observations \cite{mina_remarks_2025}. This paper extends these principles to the MARL domain, where coordination is challenged by heterogeneous inter-agent delay regimes and intermittent communication loss. We implement our estimator as an execution-time wrapper for pre-trained multi-agent policies, effectively decoupling the state-estimation process from the policy optimization objective.

\section{Problem Statement}

We model delayed multi-agent execution as a decentralized partially observable Markov decision process (Dec-POMDP), written as $\langle S, A, P, R, \Omega, O, \gamma \rangle$. At time $t$, the environment occupies a latent global state $s_t \in S$. Each agent $i$ selects an action $a_t^i$ from its local information, contributing to the joint action $\mathbf{a}_t$, which determines the transition to $s_{t+1}$ and the reward $r_t$. Because agents do not observe $s_t$ directly, each agent instead receives a partial local observation $o_t^i \in \Omega$.

In addition to local sensing, agents communicate over a channel with stochastic delay and packet loss. If agent $j$ transmits a message to agent $i$ at time $t$, the message arrives at time $t + \tau_{i,j}$, where $\tau_{i,j}$ is a random delay sampled from  $H_{i,j}(t)$. When packets are dropped, no new information from $j$ is received until communication resumes. As a result, different agents may act on views of the world that are not only partial, but also temporally misaligned.

This temporal mismatch is the core difficulty. At decision time, agent $i$ must choose an action using local observations together with communicated information that may be several steps old. Formally, the policy input is no longer a synchronized observation of the current state, but a collection of stale and possibly missing messages. The execution problem is therefore not only partially observable, but also delay-corrupted.

Our objective is to mitigate this corruption at execution time. Rather than modifying the MARL training algorithm, we seek an estimator that maps the delayed message stream available to agent $i$ into a current belief-state estimate $b_{i,t}$ that is more aligned with the underlying state relevant for control. The policy can then act on $b_{i,t}$ instead of directly consuming stale communicated observations. This is the problem solved by the execution-time filtering layer introduced in the Approach section.

\section{Approach}

\begin{figure*}[t]
    \centering
    \includegraphics[width=\textwidth]{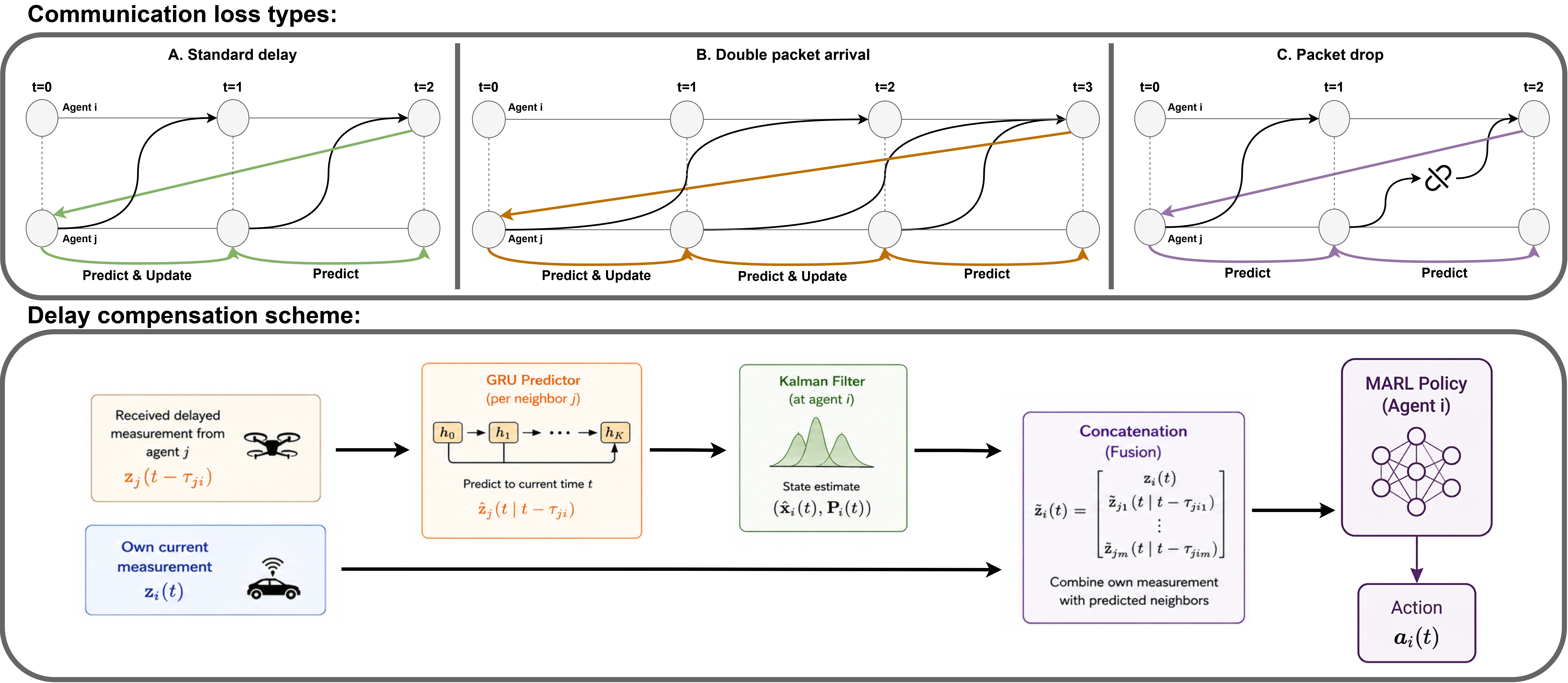}
    \caption{Illustration of the delay compensation layer. The top panel depicts three communication regimes: (A) standard latency, (B) simultaneous multi-packet arrival, and (C) packet loss. Black arrows represent the external information flow from peer agents, while colored arrows denote internal recursive state manipulations. The bottom panel details the algorithmic flow corresponding to these colored transitions, illustrating the temporal alignment of received packets and the subsequent policy execution on the synchronized belief state.}
    \label{fig:kalman method}
\end{figure*}

The design objective is to develop a modular plug-in that maps the delayed communication history $\mathcal{H}_{i,t}$ to the belief state $b_{i,t}$ required by the decentralized policy, while preserving the original MARL backbone. Unlike approaches that incorporate delay compensation into policy training, the proposed method operates exclusively at execution time. The policy generates actions based on input observations, where the communicated components are replaced by filtered estimates of the current multi-agent state. We assume that packets are timestamped, policy execution rates are fixed, and packet ordering is preserved for each sender-receiver pair. Under these assumptions, communication delay is treated as a state-estimation problem involving stale or missing measurements. Consequently, rather than providing delayed raw data to the policy, agent $i$ maintains and updates a belief over the current control-relevant states of neighboring agents.

\subsection{Belief Filtering Layer}

For each neighbor $j \neq i$, agent $i$ maintains a recursive estimate of the communicated state vector $x_{j,t}$, encompassing task-relevant features such as position, velocity, or transmitted latent variables. This representation is updated at each timestep using the most recent information available from all peers. Upon receiving a packet from agent $j$, agent $i$ performs a predict-update cycle and propagates the resulting estimate from the packet’s timestamp to the current control step. The complete delay compensation architecture is illustrated in Figure \ref{fig:kalman method}.
This framework facilitates three distinct execution modes: a standard predict-update cycle for synchronous packet arrival, a multi-update sequence for instances where multiple packets arrive within a single control interval, and an open-loop prediction mode during temporary communication failure. In every scenario, the estimator supplies the policy with a synchronized state estimate rather than a stale observation. This recursive process directly addresses the delayed-observation model defined in the problem statement. As illustrated in Figure \ref{fig:kalman method}, the module handles various communication conditions as follows:
\begin{itemize}
\item[A)] Under timely communication, the module functions as a conventional predict-update estimator.
\item[B)] If multiple packets arrive asynchronously, the estimator processes them in chronological order.
\item[C)] If communication is interrupted, the module reverts to open-loop prediction from the last verified state until new data is received.\end{itemize}
Consequently, the policy consistently operates on a synchronized belief feature instead of a stale message.

\subsection{Learned Transition Model}
To eliminate the need for hand-crafted dynamics and ensure cross-domain generalizability, we learn the transition model directly from delay-free rollout buffers collected during the initial policy training phase. Specifically, we train a Gated Recurrent Unit (GRU) \cite{chung_empirical_2014} to approximate the local dynamics of the communicated state vector. To account for the varying numerical ranges across different observation features, all inputs are normalized prior to training. Furthermore, we restrict training to data sampled from the latter stages of policy convergence, where the trajectories exhibit greater stability. Following standard residual dynamics formulations \cite{nagabandi_neural_2017}, the model is parameterized to predict a residual increment rather than the absolute next state, which enhances short-horizon stability and captures subtle transitions between consecutive timesteps:
\begin{equation}
x_{t+1} = x_t + \Delta x_t \approx x_t + \mathrm{GRU}(x_t,h_t)
\end{equation}
During training, the GRU hidden state is propagated throughout the sequence and reset only upon environment termination. At execution time, the GRU serves as a learned surrogate for the dynamics transition matrix within the estimator. Although the GRU non-linearly maps state transitions, calculating its exact Jacobian at each control step is computationally expensive for real-time robotic applications \cite{griewank_evaluating_2008}. To maintain computational efficiency, we approximate the transition Jacobian with an identity matrix. This simplification assumes a first-order persistence of uncertainty and uniform locally linear variance for covariance propagation. The policy, critic, and MARL objectives remain unmodified, preserving the modularity of the architecture and ensuring the estimator can be integrated with diverse pre-trained policies with minimal engineering overhead.

\subsection{Belief Construction and Policy Interface}
We distinguish between the predict-update cycle and the rollout. When new information is received, a predict-update cycle is performed to update the agent's inner state and the GRU's hidden state. In contrast, a rollout involves saving the most recent verified state and the GRU's hidden state then performing a prediction for the required $\tau$ steps. This saved hidden state and inner state are then used as the basis for the subsequent predict-update cycle; this prevents the state vector from being contaminated by the cumulative momentum or error drift inherent in the hidden state during an open-loop rollout.

After prediction and correction, agent $i$ aggregates its local observation, assumed to be known perfectly, with the current neighbor estimates to form the belief feature $b_{i,t}$. This belief is the only quantity exposed to the policy. Consequently, the interface remains identical to the one described in the problem statement:
\begin{equation}
a_{i,t} \sim \pi_{\theta_i}(b_{i,t}),
\end{equation}
where the communicated components of $b_{i,t}$ now correspond to current state estimates rather than stale packets. The estimator functions as middleware between the environment and the controller: it predicts through latency, incorporates asynchronous information when available, and provides a temporally aligned representation for action selection.

\begin{algorithm}[t]
\caption{Delay-aware belief update for agent $i$}
\label{alg:delay_aware_belief_update}
\begin{algorithmic}[1]
\REQUIRE $(\hat{x}_{i,t-\tau-1}, P_{i,t-\tau-1})$, $o^{\mathrm{loc}}_{i,t}$, received packets $\mathcal{M}_{j,t}$
\FORALL{packet $m_{j\to i} \in \mathcal{M}_{j,t}$}
    \IF{$\mathcal{M}_{i,t}=\emptyset$}
        \STATE $h^*_{j,t-\tau-1} \leftarrow \text{Save GRU state } h_{j,t-\tau-1}$
        \STATE $(\hat{x}_{j,t}, P_{i,t}) \leftarrow \textsc{Rollout}($
        \STATE \quad $\hat{x}_{j,t-\tau-1}, P_{j,t-\tau-1}, h_{j,t-\tau-1})$ 
    \ELSE
        \STATE Load and set GRU inner state $h^*_{j,t-\tau-1}$
        \STATE $(\hat{x}_{j,t-\tau}, P_{i,t-\tau}) \leftarrow \textsc{Predict}($
        \STATE \quad $\hat{x}_{i,t-\tau-1}, P_{i,t-\tau-1}, h^*_{j,t-\tau-1})$ 
        \STATE $\textsc{Update}(\hat{x}_{j,t-\tau}, \hat{z}_{i,t-\tau})$
        \STATE $h^*_{j,t-\tau} \leftarrow \text{Save GRU state } h_{j,t-\tau}$
        \STATE $(\hat{x}_{j,t}, P_{j,t}) \leftarrow \textsc{Rollout}(\hat{x}_{j,t-\tau}, P_{j,t-\tau}, h_{j,t-\tau})$
    \ENDIF 
\ENDFOR
\STATE $b_{i,t} \leftarrow \text{Build belief from local/neighbor estimates}$
\STATE Select action $a_{i,t} \sim \pi_{\theta_i}(b_{i,t})$
\end{algorithmic}
\end{algorithm}

As detailed in Algorithm \ref{alg:delay_aware_belief_update}, this design offers several practical advantages. First, the belief representation size remains constant, avoiding the input dimensionality growth associated with long history windows. Second, the recursive prediction naturally accommodates variable delay patterns, including bursty latency and intermittent packet loss. Third, the system maintains functionality during communication failures by reverting to open-loop propagation. Finally, because the estimator is decoupled from policy optimization, it is agnostic to the underlying MARL backbone and can be deployed as a wrapper for existing controllers.

\section{Experiments}

\begin{figure*}
    \centering
    \includegraphics[width=0.16\linewidth]{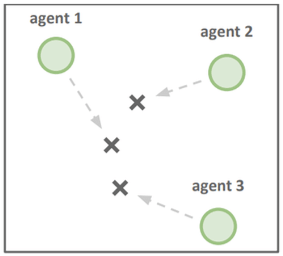}\hfill
    \includegraphics[width=0.16\linewidth]{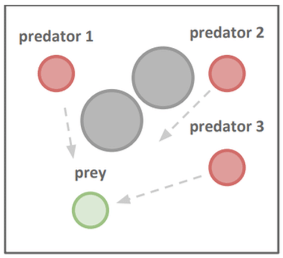}\hfill
    \includegraphics[width=0.16\linewidth]{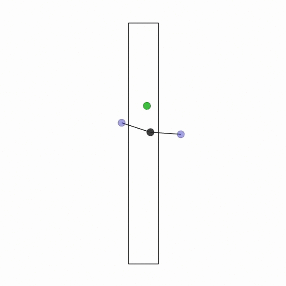}\hfill
    \includegraphics[width=0.16\linewidth]{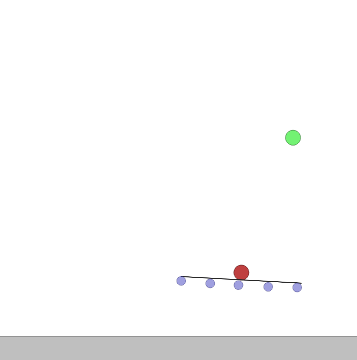}\hfill
    \includegraphics[width=0.16\linewidth]{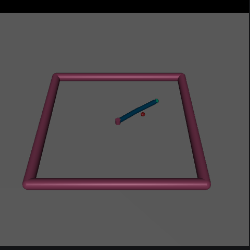}\hfill
    \includegraphics[width=0.16\linewidth]{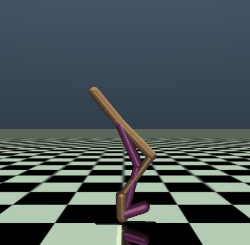}
    \caption{Evaluation benchmarks (ordered left to right): MPE Spread and Tag, involving landmark navigation and prey interception; VMAS Buzz-wire and Balance, requiring cooperative manipulation of a shared object toward a target; and Multi-Agent Reacher and Walker2d, focusing on multi-link robotic arm control and coordinated bipedal locomotion.}
    \label{fig:envs}
\end{figure*}

We evaluate the proposed execution estimation layer across six environments spanning varying levels of dynamical complexity, observability, and coordination requirements (Figure \ref{fig:envs}). The experimental framework is designed to validate three primary claims: (i) the method generalizes across diverse MARL backbones and benchmark families; (ii) the learned GRU dynamics model is sufficiently accurate to support recursive filtering in both simplified and high-fidelity domains; and (iii) the resulting execution state estimates remain effective under increasing and out-of-distribution (OOD) delay patterns. To evaluate these conditions, we synthetically inject communication delays between the ground-truth observations generated by the environment and the observations received by the agents.

In MPE \cite{lowe_multi-agent_2020}, we compare our approach against the RDC pipeline using the open-source implementation based on QMIX \cite{rashid_qmix_2018}. In VMAS \cite{bettini_vmas_2022}, we evaluate more complex interactions involving rich continuous dynamics and partially observable agents. Further evaluation is conducted in high-fidelity continuous-control settings using MaMuJoCo \cite{peng_facmac_2021}. For both continuous-control tasks, we employ MAPPO-style policies \cite{yu_surprising_2022} and report results across 8 random seeds with 50 evaluation episodes per seed. Across all domains, we compare the method to a delay-unaware baseline; in the MPE environment, we additionally compare against the RDC framework.

Unless otherwise specified, Kalman gains are not tuned per environment. The process covariance is set based on the GRU training MSE error for each state component, and the observation covariance is defined as a fixed relative scaling of that value. This simplified parameterization is intended to demonstrate the method's efficacy without the need for extensive estimator-specific hyperparameter optimization. We do not claim formal optimality for the estimator; the empirical objective is to determine whether a lightweight execution filter is sufficient to recover robustness under varying delays.

\subsection{World Model Convergence and Fidelity}
\begin{figure}
    \centering
    \includegraphics[width=\linewidth]{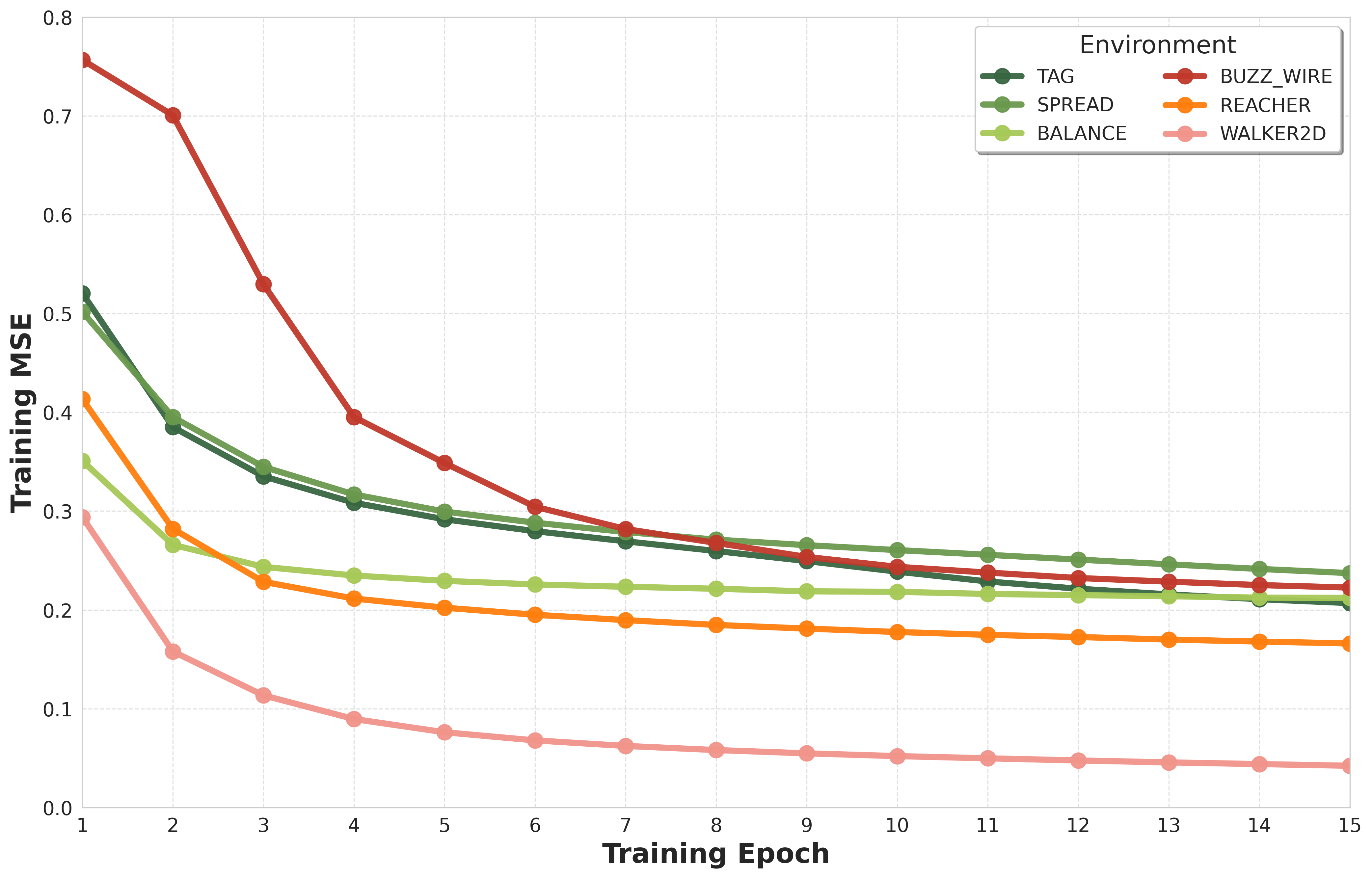}
    \caption{Normalized MSE convergence for the GRU dynamics model across all benchmark environments over training epochs.}
    \label{fig:training_mse}
\end{figure}

We first analyze the learned dynamics model. Two key observations were critical for training stability. First, expressing positions in absolute coordinates simplified the prediction task by mitigating relative-motion ambiguity. Second, high-fidelity simulators required actuation smoothing to ensure a sufficient signal-to-noise ratio; otherwise, the model tended to fit high-frequency noise rather than the underlying dynamics. With this preprocessing, all GRU models converged rapidly, typically within 15 epochs (Figure \ref{fig:training_mse}).

The final prediction quality varies systematically with environment structure. Walker2D and Reacher, which were trained with smoother policies, achieved the lowest training errors. This suggests their dynamics are sufficiently regular for the GRU to learn precise short-horizon transition models once high-frequency control jitter is reduced. Conversely, Tag and Spread exhibited a higher irreducible error floor (approximately 0.20), consistent with the fact that multi-agent intent introduces stochasticity not directly predictable from local state-action pairs. This effect is further amplified by discrete action spaces and significant time quantization, which induce larger transition gaps between states. Nevertheless, across all domains, the learned world models capture the majority of the deterministic transition manifold, even in environments with higher uncertainty.

\subsection{Cross-Domain Generalization and Resilience}
\begin{figure*}[!t]
    \centering
    \includegraphics[width=\textwidth]{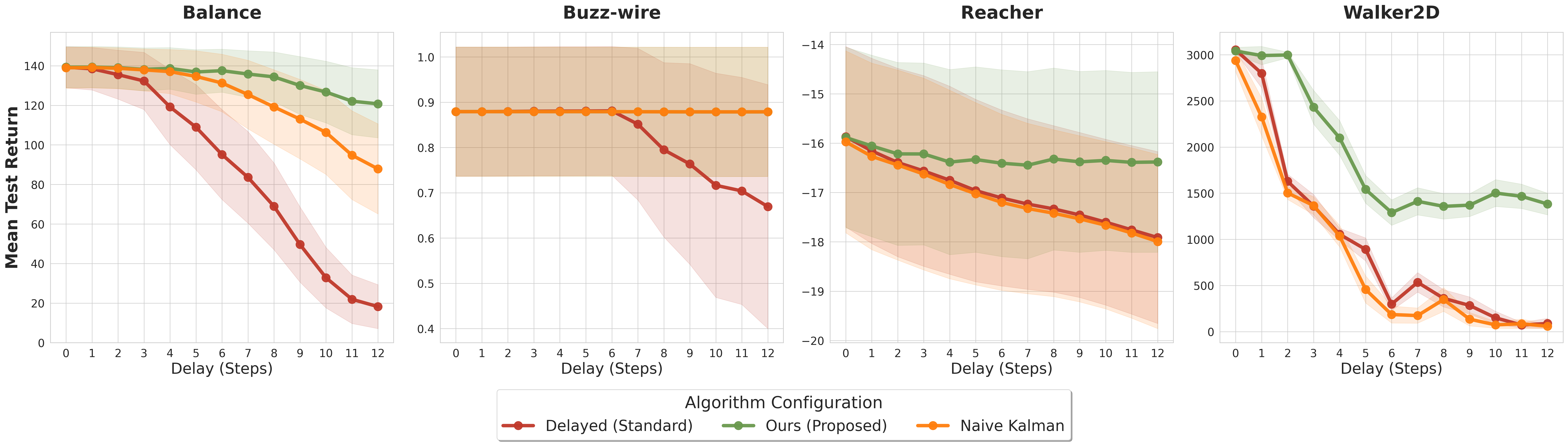}
    \caption{Robustness analysis across multiple environments: Mean and STD Performance comparison between the proposed estimation layer, Naive kalman implementation, and a baseline with no delay compensation across varying communication delays.}
    \label{fig:cross_domain_resilience}
\end{figure*}

Cross-domain experiments demonstrate that the proposed filtering architecture generalizes across a diverse suite of control tasks. As illustrated in Fig. \ref{fig:cross_domain_resilience}, the performance of standard delayed execution degrades rapidly with increasing latency, whereas the proposed predictive layer maintains significantly higher stability. To isolate the importance of the learned transition model, we perform an ablation study comparing our approach to a damped first-order kinematic Kalman filter (Naive Kalman). This baseline was selected to ensure a fair comparison regarding generalization, avoiding the need for hand-coded, task-specific dynamics models. The Naive Kalman is initialized using the same uncertainty parameters as the GRU-based variant.

In the Reacher and Buzz Wire environments, performance gains are primarily driven by enhanced precision; the filter mitigates the overshoot and synchronization errors that typically result from stale state feedback. The impact is most pronounced in Walker2D—the most challenging case study due to its high-dimensional coordination requirements. In this task, the delayed baseline collapses even at minimal latency, while the filtered policy maintains a stable bipedal gait across a much broader delay range.

When compared to the Naive Kalman, we observe that the kinematic baseline fails as expected in the Reacher and Walker2D environments, where the dynamics are highly coupled and non-linear. In the Balance task, the Naive Kalman provides a slight improvement, though it fails to match the performance of the GRU-based model, indicating that while basic estimation helps, it cannot fully capture the environment's underlying dynamics. In the Buzz Wire task, both estimation models perform similarly; this is consistent with the environment’s relatively simple kinematic structure, which is well-approximated by a first-order model. These results indicate that our method is effective beyond simple kinematic tasks. Even for articulated systems with highly non-linear dynamics, the GRU-based transition model generates sufficiently accurate short-horizon predictions to enable effective Kalman-style latency compensation.

\begin{figure}[H]
\centering
\includegraphics[width=\linewidth]{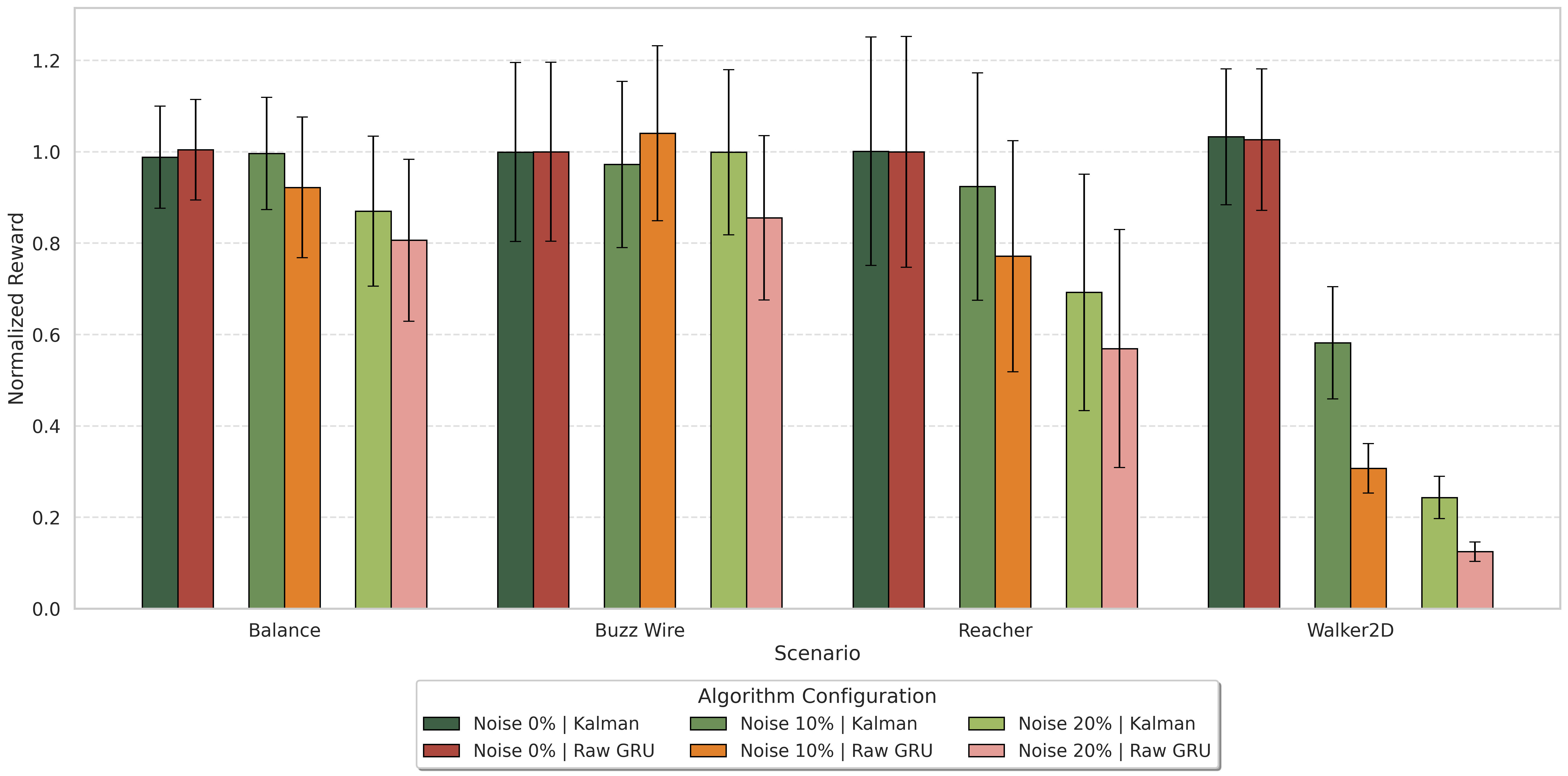}
\caption{Robustness to observation noise: Normalized reward performance comparing the proposed Kalman-filtered layer against a standalone GRU transition model. The results are evaluated across four environments under three distinct noise injection regimes.}
\label{fig:NoiseResillence}
\end{figure}

To isolate the contribution of the recursive filter to noise robustness, we compare the Kalman-wrapped GRU against a standalone GRU predictor (Fig. \ref{fig:NoiseResillence}). In this ablation, the standalone model performs open-loop rollouts from the last known observation, whereas the filtering layer recursively assimilates noisy measurements. This setup evaluates the framework’s efficacy in real-world scenarios characterized by stochastic sensor error. As illustrated in Fig. \ref{fig:NoiseResillence}, the benefits of the recursive update scale proportionally with observation uncertainty. Under zero-noise conditions, the Kalman-based and standalone GRU models achieve comparable performance, as both rely on accurate initial states. However, as the noise amplitude increases, the filtering layer demonstrates superior resilience. This is most prominent in the Walker2D environment, where the filtered policy maintains significantly higher rewards at 10\% and 20\% noise levels. Such results suggest that for delay-sensitive and dynamically unstable systems, robust state estimation is indispensable for control stability. While the Reacher and Balance tasks show more marginal gains, they maintain a consistent advantage over the standalone baseline, further showcasing the utility of the filtering layer alongside inherent benefits such as multi-sensor fusion and explicit uncertainty estimation.

The Buzz Wire environment represents a notable outlier: the performance gap there is negligible and slightly favors the standalone GRU at 10\% noise. This suggests that the policy for this task is either less sensitive to high-frequency noise or the environment dynamics are inherently more robust to minor state perturbations. These findings indicate that the Kalman wrapper is most effective in settings where delayed execution amplifies minor state-estimation errors into significant control failures. The effect is most pronounced in unstable or tightly coupled domains and less significant in tasks with higher error tolerance. This alignment with traditional Kalman filter applications underscores the layer's utility for real-world deployment, offering structural advantages like sensor fusion and covariance estimation that are absent in standalone recurrent models.

A primary practical advantage of this method is its utility as a modular execution layer. A policy trained in a delay-free environment can be adapted to delayed settings by inserting the estimator between the environment and the agent, effectively providing a "zero-shot" transition to non-ideal hardware. This modularity obviates the need for retraining delay-aware controllers from scratch, making it an ideal candidate for legacy systems or proprietary control architectures.

\subsection{Comparative Analysis in MPE}
\begin{figure}
\centering
\includegraphics[width=\linewidth]{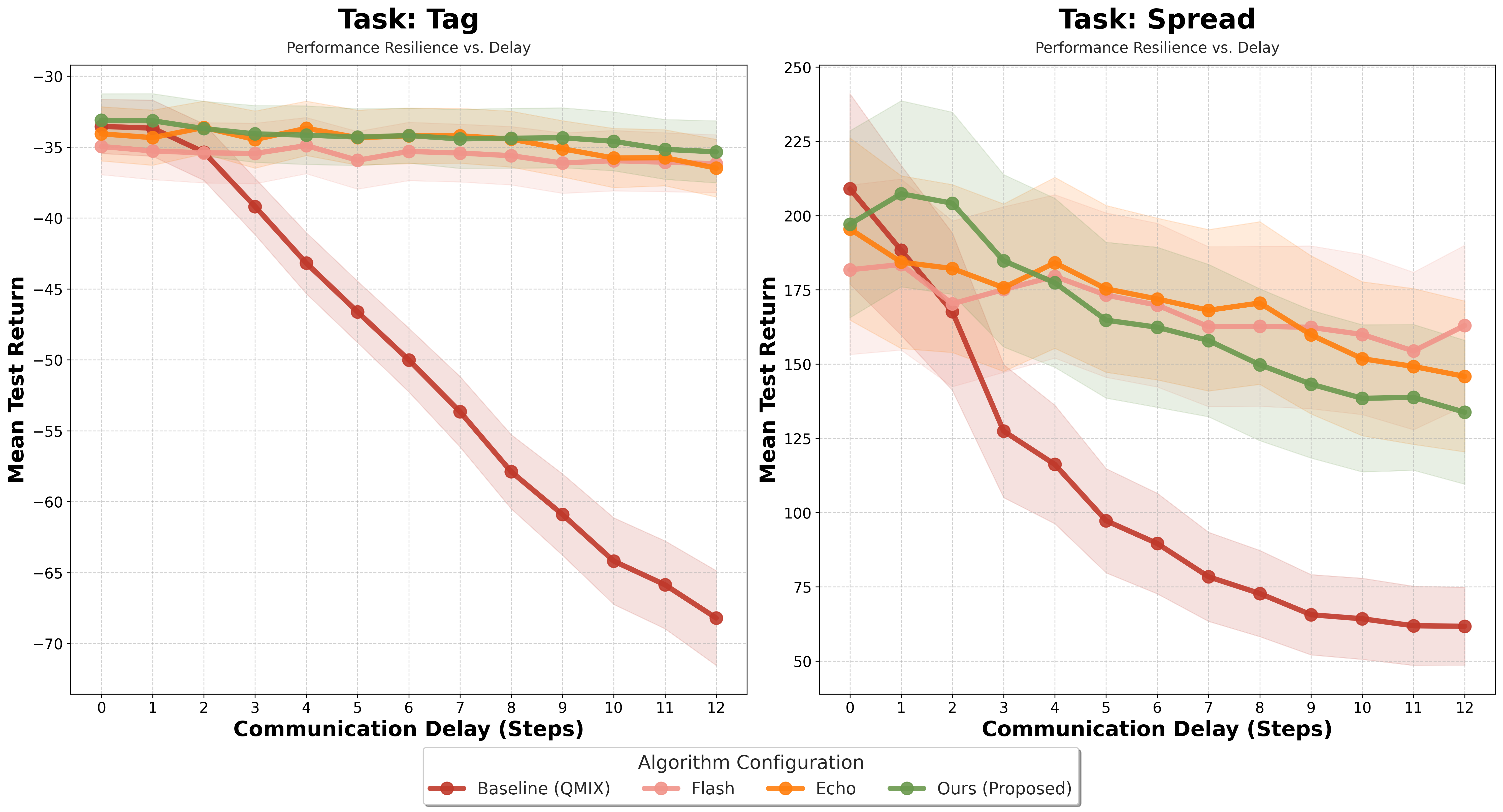}
\caption{Robustness analysis on MPE Spread and Tag: Mean and STD Performance comparison between the proposed estimation layer, Rainbow Delay Compensation (RDC), and a baseline with no delay compensation across varying communication delays.}
\label{fig:mpe_resilience}
\end{figure}

In the MPE experiments, we conduct two primary comparisons: mean return as a function of increasing latency and resilience to varying delay windows, following the evaluation protocol established in the RDC framework. As illustrated in Fig. \ref{fig:mpe_resilience}, our approach demonstrates significant performance gains over the baseline in both tasks. These results are comparable to the RDC framework, further confirming the generalizability of our belief-state layer across diverse multi-agent environments.

\begin{figure}[H]
\centering
\includegraphics[width=\linewidth]{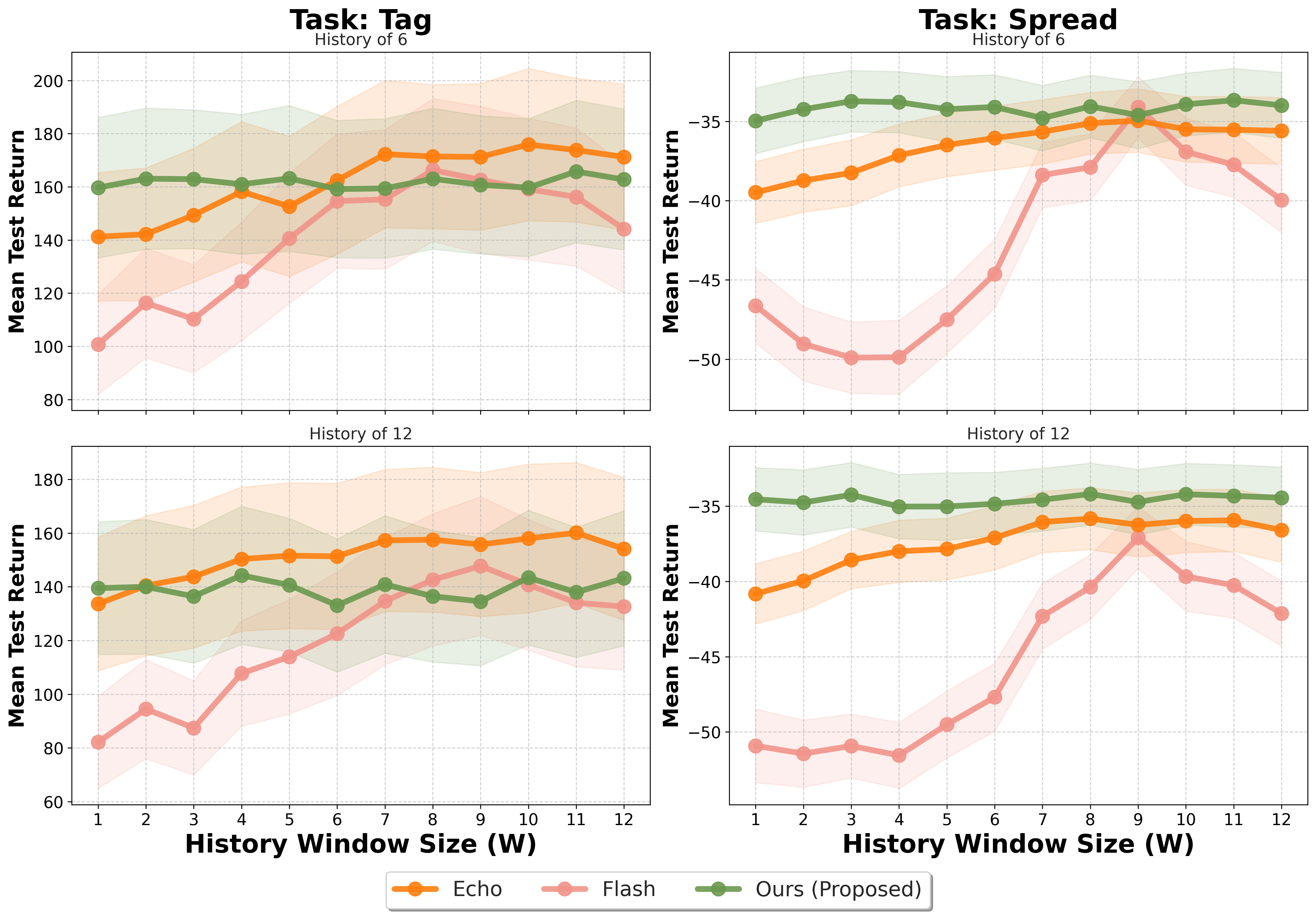}
\caption{Impact of history-window size on reward performance: Mean and STD Comparison between the proposed method, Rainbow Delay Compensation (RDC), and an uncompensated baseline under fixed latencies of 6 and 12 steps in MPE environments.}
\label{fig:history_window}
\end{figure}

Performance remains consistent across both scenarios, though returns in the Tag task are slightly lower; this is expected given the higher GRU prediction error observed in this environment. A critical distinction between our framework and methods such as FLASH or ECHO is that our architecture is entirely decoupled from policy training. While those methods achieve robustness via learned delay-compensation modules integrated into the RL backbone, often requiring extensive retraining and showing sensitivity to history-window design, our approach is modular. By combining a learned transition model with a fixed recursive update, our execution layer facilitates seamless integration with pre-trained policies and reduces sensitivity to temporal-window hyperparameters. This advantage is clearly demonstrated in the history-window analysis (Fig. \ref{fig:history_window}).

Our method exhibits a stable response across various window sizes, indicating that the recursive filtering-prediction loop effectively propagates state information without relying on a brittle, fixed-length memory buffer. This hyperparameter robustness is practically significant, as it eliminates the need for environment-specific retraining when deployment delays differ from those encountered during training. Furthermore, the method maintains consistent performance under out-of-distribution (OOD) delay patterns, scenarios where integrated agent-estimator models typically degrade if the deployment window does not match the training configuration.

The computational overhead of the proposed framework is modest. For a six-step delay, the recursive filtering step requires approximately 0.007 s when tested on hardware comparable to that reported in the original RDC study. While cross-paper runtime comparisons should be interpreted with caution, this latency is comparable to the FLASH model and significantly lower than the ECHO model, all while maintaining competitive performance against more complex delay-compensation baselines.

\section{Conclusion}
We propose a modular execution-layer state estimator for multi-agent reinforcement learning (MARL) under delayed communication. Rather than introducing a new MARL training algorithm, we compensate for stochastic communication delays and packet loss by inserting a belief-state estimator between the environment and a pre-trained policy. This execution layer integrates a learned GRU transition model with a Kalman recursive update to replace stale observations with current state estimates.

Empirical evaluation demonstrates that this execution layer enhances robustness across MPE, VMAS, and high-fidelity continuous-control benchmarks. Three primary conclusions emerge: (i) the learned GRU dynamics are sufficiently accurate to support recursive filtering across heterogeneous tasks; (ii) the resulting belief updates remain effective across a broad spectrum of delay magnitudes; and (iii) the method is resilient to observation noise without requiring environment-specific history buffer tuning. The most significant performance gains occur in highly delay-sensitive domains, where acting on stale feedback otherwise leads to rapid control instability.

While effective as a modular execution-layer compensator, the current framework does not claim to be a globally optimal estimator. Future research could investigate analytical refinements to the filtering logic or algorithmic optimizations to further tighten estimation bounds. Another promising avenue involves the joint training of the policy and estimator. By exposing the policy to the estimator’s internal uncertainty metrics, such as the predicted covariance, agents could learn to adopt risk-aware or cautious control strategies during periods of high uncertainty. Such a tightly-coupled approach could significantly enhance robustness in environments characterized by severe model mismatch or highly non-linear multi-agent interactions.

\section*{Acknowledgments}
The authors acknowledge the financial support provided by
the University of Haifa
 
\bibliographystyle{unsrt}
\bibliography{references}

\vfill

\end{document}